\documentclass[aps,prl,twocolumn,10 pt,superscriptaddress,floats,showpacs]{revtex4}
\usepackage[dvips]{graphicx}
\begin{document}

\title{quantitative analysis of the critical current due to vortex pinning by surface corrugation.}
\author{A. Pautrat}
\email[corresponding author: ]{ alain.pautrat@ismra.fr} \author{J. Scola}\author{C.
Goupil}\author{Ch. Simon} \affiliation{CRISMAT/ENSICAEN, UMR 6508 du CNRS,14050 Caen,
France.}
\author{C. Villard}
\affiliation{CRTBT/CRETA, UPR 5001 du CNRS, 38042 Grenoble, France.}
\author{B. Domeng$\grave{e}$s}
\affiliation{LAMIP, Phillips/ENSICAEN, 14050 Caen, France.}
\author{Y. Simon}
\affiliation{LPMC de l'Ecole Normale Sup\'erieure, UMR 8551 du CNRS, associ\'ee aux
universit\'es Paris 6 et 7, 75231 Paris Cedex5, France.}
\author{C. Guilpin}
\affiliation{GPS, UMR 7588 du CNRS, associ\'ee aux universit\'es Paris 6 et 7, 75251
Paris Cedex5, France.}
\author{L. M$\acute{e}$chin}
\affiliation{GREYC/ENSICAEN, UMR 6072 du CNRS,14050 Caen, France.}

\begin{abstract}
The transport critical current of a Niobium (Nb) thick film has been measured for a large
range of magnetic field. Its value and variation are quantitatively described in the
framework of the pinning of vortices due to boundary conditions at the rough surface,
with a contact angle well explained by the spectral analysis of the surface roughness.
Increasing the surface roughness using a Focused Ion Beam results also in an increase of
the superficial critical current.
\end{abstract}


\newpage
\maketitle

\section{Introduction}

The understanding of the nature of the critical current in superconductors has triggered
a lot of work for many years \cite{Campbell, Pan}. Practically speaking, quantitative
predictions of the critical current values is a key point for most of the applications.
More fundamentally, the question of the interaction of the vortex lattice with pinning
centers is also a prototype system for understanding the behaviour of an elastic medium
submitted to a disorder potential \cite{Giam}. It is well known that a perfect vortex
lattice (VL) submitted to a bulk transport current is subject to a flow, that leads to a
finite electro chemical field and gives rise to dissipation \cite{Thorel}. The ability of
the VL not to move when it is submitted to a transport current is generally explained by
the pinning interactions between pinning centers and VL. Unfortunately, the exact nature
of the pertinent pinning centers, and the way they are acting, is  not straightforward.
In soft samples where superconducting parameters vary slowly in the sample,
cristallographic bulk inhomogeneities are usually supposed to play the role of pinning
centers \cite{Pan}. One can notice that this collective bulk pinning description leads to
quite complex critical current expressions and to a corresponding lack of quantitative
interpretation and prediction. On the other hand, experimental observation of the strong
influence of surface quality on VL pinning has been evidenced in pioneering experiments
\cite{surf}. This surface influence can be described in the framework of surface
roughness interacting with VL, as proposed by Mathieu-Simon (MS) in a continuum approach.
For any real sample, which presents a surface roughness at the VL scale, the respect of
the boundary conditions \cite{notabene} of terminating vortices at the surface imposes
local bending. This defines a contact angle $\theta$ ($\theta$=0 for a flat surface) and
leads to a near-surface supercurrent. At the sample scale, this offers a lot of
metastable equilibrium states. The ability of the system to sustain a macroscopic
supercurrent is then directly linked to the surface roughness via an average contact
angle $\theta_c$. The macroscopic critical current (per unit of width) is simply given by

\begin{equation}
  i_{c} (A/m) = \varepsilon.sin(\theta_{c})
\end{equation}

where $\varepsilon$ is the vortex potential, i.e. the thermodynamical potential
describing the equilibrium state. "Pinning" is then nothing else than the consequence of
vortex boundary conditions applied to a real surface \cite{gilles}. This should a priori
act in every sample. The remaining question is: what is the contribution of this surface
critical current with respect to the overall critical current? The vortex potential can
be measured from the reversible magnetic moment M = -$\int_{V}\varepsilon$ $\delta$V of a
superconducting volume V, or calculated using, for example, Abrikosov's solution close to
$B_{c2}$. Using an iteration procedure proposed by E.H. Brandt \cite{mumut},
$\varepsilon$ can be also obtained over the whole mixed state via numerical solutions of
the Ginzburg-Landau equations. Now if one assumes a physical value of $\theta_{c}$ = 0-20
deg, a good order of magnitude of $i_{c}$ is deduced \cite{placais}. A careful inspection
of the surface state should in principle enable to extract a more precise value of this
critical angle. If the usual measurement of surface roughness is the rms
(root-mean-square) height $h$ of the surface bumps, the pertinent parameter is here more
the distribution $\theta(x)$ = $\arctan$ ($\frac{dh}{dx})$ of local slopes over the width
of the surface. The rms value of this angle is given by the integration of the spectral
density $\overline{\theta^{2}}= \int_{k_{min}}^{k_{max}} S_{\theta\theta}$ $dk$ over the
appropriate $k$ boundaries \cite{landau}. $S_{\theta\theta}$ is the Fourier transform of
the autocorrelation of $\theta(x)$ (Wiener-Khintchine theorem), which represents the
spatial distribution of $\theta$.
\smallskip
The aim of the present study is to measure the critical current of a Niobium thick film,
with various surface corrugation, and to compare the experimental values to those
obtained with the simple expression (1). $\varepsilon$ will be calculated using Brandt's
approach, and $\theta_c$ compared to $\sqrt{\overline{\theta^{2}}}$.

\section{Experimental}

The sample used is a film of Niobium (thickness= 3000 $\AA$) deposited at 780°C on a
sapphire substrate by the ion beam technique. The film has a resistivity of about 0.5
$\mu\Omega$.cm at the critical temperature $T_{c}$ = 9.15 K and exhibits a low surface
rms roughness ($R_{a}$ $\lesssim$ 5 nm), measured by Atomic Force Microscopy (Nanoscope
III, Digital Instruments). Microbridges of W=10 $\mu$m $\times$ L=30 $\mu$m have been
patterned using a scanning electronic microscope, this irradiation step being followed by
a reactive ion etching process. The critical currents have been measured by mean of the
standard four-probe technique, at the following temperatures of 4.2, 5.2, 6.2 K in the
whole range of field covering the mixed state. The critical current values $I_{c}$ were
determined with a voltage criteria of 10 nV.

\section{Results and discussion}

Let us first discuss the general behavior of the $I_{c}$($\omega$) data (fig.1) for the
virgin microbridge at different temperatures ($\omega$ is the vortex field inside the
sample and corresponds to the vortex density n=$\frac{\omega}{\phi_{o}}$) . Note that the
demagnetization factor due to the geometry of the thin film renormalizes the apparent
first critical field $B_{c1}$ up to about 0.015 $B_{c1 bulk}$. It implies that the mixed
state is created for the lowest field value applied ($\approx 30 Gauss$). Except for this
peculiarity , $I_{c}$($\omega$) exhibits the same field variation as the reversible
magnetization curve of a type II superconductor, in agreement with the expected variation
of superficial currents $i_{c}$ (equation 1 with $\theta_{c} \approx$ cte). We also
notice that the curves taken at 4.2, 5.2 or 6.2 K are self similar, that is they can be
superimposed by a mere rescaling. As first noted in pioneering work on vortex pinning
\cite{archie}, the change of critical current with temperature can be totally attributed
to the change in primary superconducting properties. It is evidenced in figure 2 where
the low field value of $I_{c}$ is shown to be simply proportional to $B_{c2}$ for the
three temperatures and both for the virgin and for the damaged sample. This shows that
the variation of $I_{c}$ with temperature is simply due to the variation of the vortex
potential $\varepsilon$(T)(i.e. the variation of superfluid density), without the need of
involving other thermal effects such as vortex thermal diffusion. Now if we want to
verify quantitatively equation (1), we first need to know the vortex potential
$\varepsilon$. It is usually approximated using the Abrikosov calculations from B$_{c2}$
down to 0.4-0.5 $B_{c2}$, and by the London expression at very low fields $B \gtrsim
B_{c1}$. For low kappa superconductors such as pure Niobium ($\kappa\approx1$ for a
sample with properties very close to ours \cite{Auer}\cite{Weber}), only the Abrikosov
expression is quantitatively correct and it is necessary to use numerical calculations to
solve the Ginzburg Landau equations over the whole field range, following for example
Brandt's iterative method \cite{notabene2}. The result is presented in figure 3, for
$\kappa$ = 1. We can now deduce the critical angle needed to account for the measured
critical current. For a pure surface pinning and following equation (1), the expected
value of $\theta_{c}$ is given by $\theta_{c}$ $\approx$ arcsin ($i_{c}$/$\varepsilon$).
Using the numerical values of $\varepsilon$, we find $\theta_{c}$ $\approx$ 0.4-1 deg
(see figure 4), in agreement with the mere expectation of a physical angle. The order of
magnitude is promising, but the complexity of the (multiscale) surface disorder needs a
careful surface analysis. We have therefore measured the microbridge roughness using AFM
in tapping mode (see fig. 5).

 Following the simple analysis described in the introduction, we obtain the spectral density
$S_{\theta\theta}$ with the use of the Fourier transform of the autocorrelation of
$\theta(x)$ (Wiener-Khintchine theorem). The main value of the statistically
representative angle is given by the integration of the spectral density \cite{landau}:
$\overline{\theta^{2}}= \int_{0}^{k_{max}} S_{\theta\theta} dk$. The boundary of the
integral have been chosen as the natural scale for the vortex lattice, considering that a
vortex does not see a roughness less than its core size ($k_{max}=\frac{2\pi}{\xi}$).
Note that the choice of this cutoff frequency does not significantly change the results.
This calculation gives $\sqrt{\overline{\theta^{2}}}$ $\approx$ 0.60 $\pm$ 0.18 deg. The
 agreement with the value deduced from the critical current measurements (a mean value of 0.70 $\pm$ 0.15 deg) is
 quite promising. More precisely, this is, within error bars, what we obtained for the magnetic field values
 higher than about 1000 Gauss. It is worth noting that the main value that we calculate is statistically representative
  but gives also a value that is supposed to be independent of the frequency. We are fully aware that a more rigorous analysis should take into account a kind of
matching effect between the VL periodicity and the scale of surface disorder. It is even
possible to expect a peculiar variation of $i_{c}(\omega)$ in the case of a very rough
surface at a restricted spatial scale. In this respect, one can see that the highest
angles observed for low
 fields B $\lesssim$ 0.1 T (a$_{0} \gtrsim$ 0.5 $\mu$m) are quite consistent with the highest angles observed in
  the surface profile for
 a periodicity of about 1 $\mu$m.

We decided also to compare this virgin microbridge with one whose surface structure was
modified. The idea was to use a Focused Ion Beam to etch its surface following a
controlled geometry. The expected shape was that of "corrugated iron" with 12 $\mu$m by
0.1 $\mu$m trenches regularly spaced by 1 $\mu$m. Also, the etched depth should be high
compared to the initial roughness of the surface, that is 30nm here. Several attempts
have been performed using different Ga ion doses, which should be as low as possible in
order to minimize the effect of Ga irradiation, leading to the best control of the etched
surface. Thus, the final procedure was eight identical patterns (12$\mu$m x 0.10$\mu$m x
0.03$\mu$m), etched parallel using an ion current of  4 pA corresponding to a total ion
dose about 150 pC/$\mu$m$^{2}$. The sample was tilted by 45 degrees and the magnification
used was 25kX. Figure 6 is a SEM image of the resulting etched microbridge; the trenches
are evenly spaced and they yield geometric parameters close to those expected (width
about 0.15 $ \mu$m). Such low energy irradiation leads also to an implantation of
Ga$^{+}$ ions, but simulations using Monte-Carlo calculations \cite{ziegler} indicate
that it affects only a range of no more than 100 $\AA$. Furthermore, we observe neither
any change in the critical temperature ($T_{c}=9.15 K$), nor in the normal state
resistivity ($\rho_{n} \approx 0.50 \mu\Omega.cm$) and critical fields within
experimental accuracy. In order to evaluate more precisely the influence of the Ga
irradiation, another microbridge was etched using a single rectangular pattern (12µm x
8µm x 0.03µm), that is covering the whole width of the microbridge. The sample was tilted
of 45° and the total ion dose close to 190pC/$\mu$m$^{2}$. Critical current and
resistivity measurements were performed on this microbridge and were compared to those on
the virgin microbridge. We observed the same properties and specially that the critical
current is very similar (within few percents) in the two cases. This confirmed that the
Ga irradiation had no influence on the bulk physical properties of the film. The FIB
treatment, contrary to highest energy irradiation, has not modified the bulk crystalline
lattice of the material and that the important modification is in the surface structure.
We can therefore use the term "surface" damages. The main obvious result is the increase
of the critical current, as shown in fig. 7. Following the same procedure that we
performed for the virgin microbridge, one obtains $\sqrt{\overline{\theta^{2}}}$
$\approx$ 2.2 $\pm$ 0.3 deg in good enough agreement with the 1.5 $\pm$ 0.2 deg deduced
from the critical current values. We note that the kind of treatment we performed leads
to an increase of the roughness for a periodicity of about 0.1-1 $\mu$m. In the inset of
figure 7, one can evidence that the critical angle is almost unchanged for the highest
magnetic field values $\omega \gtrsim$ 3000 G. For those values, the intervortex distance
is less than 0.1 $\mu$m and we find using the AFM that the treated microbridge exhibits
the same kind of roughness as the virgin microbridge for this periodicity. Again, the
spatial dependence of the surface roughness is certainly linked with the exact
$\theta_{c}(\omega)$ variation, with $\omega$ fixing the spatial scale of the pertinent
surface disorder. More work is needed to fully describe this problem.

Finally, we conclude from this analysis that the critical current of our sample is given
by  equation (1) with a good agreement, for the virgin and for the surface damaged
microbridges. We conclude that taking into account the surface defects as main sources of
pinning enables to explain the experimental critical current values. Expressed in the
form of a surface critical current and due to the small surface corrugation of the Nb
Film (a roughness of few nm rms), the critical current appears to have standard and even
relatively small values for this low kappa superconductor ($i_{c} \approx$ 10-30 A/cm at
low fields). If one expresses the critical current in the form of a density as it is
usually made, this leads to a high value ($J_{c} \approx $ $0.5-1.5$ $10^{6} A/cm^{2}$).
We emphasize that this notion of density is by definition not justified in the case of a
current flowing under the surface and not uniformly in the bulk. As an example, Niobium
crystals with a thickness of 7.6 10$^{-3}$ cm exhibit a critical current density of
roughly 3.10$^{3}$ A/cm$^{2}$ at 0.1 T and 4.2 K \cite{desorbo}. This gives $i_{c}$=13.1
A/cm compared to 12.5 A/cm (our thin film) under the same ($\omega$,T) conditions. So,
The surface critical current is  almost the same. It follows that the difference of
thickness makes this apparent (but not physically significant in terms of pinning
"force") difference in the critical current density values. Note that the same remark
apply to other types of clean superconductors \cite{joiner}.

 Now, if one increases a lot
the number of bulk defects to obtain a spacing say less than the intervortex distance, we
can obtain a so-called hard superconductor. In this case, a bulk subcritical current can
flow by a percolating-like behavior. Anyway, in this case, the critical current density
was shown, in a lot of cases, to be proportional both to the difference in equilibrium
magnetization across the internal boundaries and to the area of internal interface per
unit volume \cite{evetts,dew,murakami}. It is worth noting that this is the same kind of
pinning by surface interactions, but in this case with internal interfaces.

 In summary, we have observed that the value of the critical current of a thick
film of a standard type II superconductor is quantitatively  explained with the vortex
pinning by surface roughness. We have also observed the increase of this critical current
caused by an increase of the surface corrugation. Furthermore, it gives a simple
explanation for the high critical current density observed in this kind of clean thin
films, compared to the moderated one observed in (thick) bulk crystals. We hope that it
gives also evidence that the interaction between the surface corrugation and the vortex
elasticity is a key point for the understanding of vortex lattice pinning and dynamics.

\smallskip
Acknowledgments: J. Scola acknowledges support from "la region Basse Normandie".

\newpage

\begin{figure}
\caption{The critical current of the virgin niobium microbridge as function of the vortex
field for three different temperatures.}
\end{figure}

\begin{figure}
\caption{The critical current at 30 Gauss as function of the second critical field for
the three temperatures and the virgin and surface damaged microbridges.}
\end{figure}

\begin{figure}
\caption{The variation of the vortex potential (or equilibrium magnetization) as function
of the magnetic field for $\kappa$=1 calculated with Ginzburg-Landau equations and
following the method explained in \cite{mumut}. The two are respectively normalized over
$H_{C2}$ and $B_{C2}$. The dashed line is the Abrikosov line.}
\end{figure}

\begin{figure}
\caption{The variation of the critical angle $\theta_{c}$ deduced from the
$\arcsin$($\frac{i_{c}}{\varepsilon}$). The straight line corresponds to the main value
extracted from the surface analysis.}
\end{figure}

\begin{figure}
\caption{AFM picture of the surface roughness for the virgin microbridge. In the inset is
shown the corresponding variation of local slopes $\theta$ over the width of the sample.}
\end{figure}

\begin{figure}
 \caption{a/ SEM image of the treated microbridge showing the eight trenches (arrowed) in the
niobium film.
    b/ Detail of one trench showing its geometric characteristics.}
\end{figure}

\begin{figure}
\caption{comparison of the critical currents for the virgin microbridge and for the one
degraded with the FIB. In the inset is shown the critical angle variation deduced from
equation (1).}
\end{figure}


\begin{thebibliography}{}
\bibitem{notabene} This boundary condition can be derived from the Ginzburg-Landau equations. It has
been also demonstrated using thermodynamic arguments in \cite{MS},and can be described in
terms of supercurrent flowing tangentially to the surface.
\bibitem{notabene2}The code has been written in C by Christian Guilpin and Yvan Simon, following the method proposed by E.H. Brandt in \cite{mumut}.
\end{thebibliography}
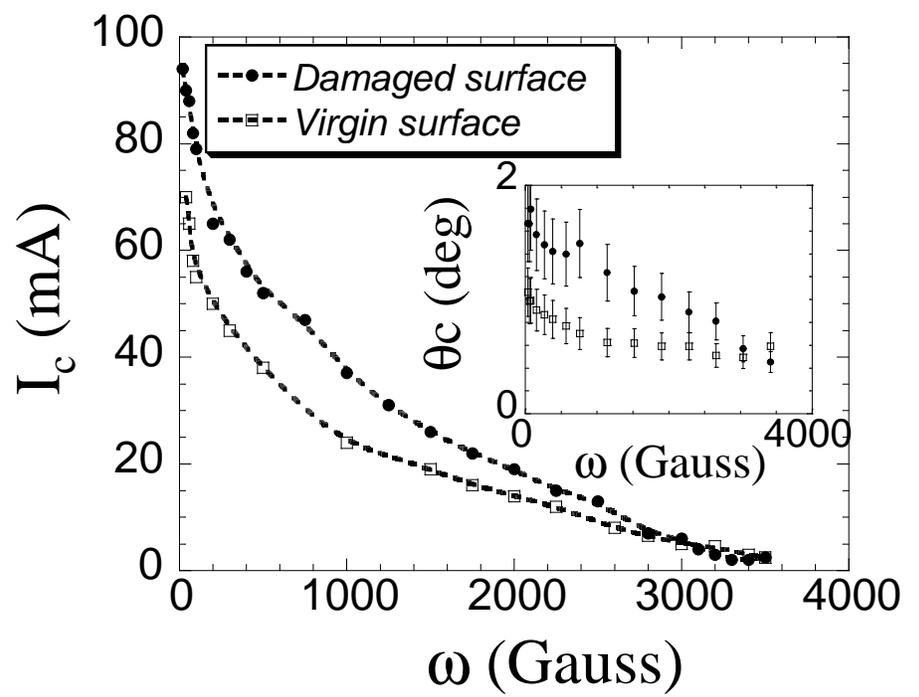
\end{document}